\documentclass[aps,groupedaddress,twocolumn]{revtex4}
\usepackage{amsfonts}
\usepackage{amsmath}
\usepackage{mathrsfs}
\usepackage{amssymb}
\usepackage{graphicx}
\usepackage{dcolumn}
\usepackage{bm}
\usepackage{color}

\begin{document}

\title{Exceptional Points in Random-Defect Phonon Lasers}

\author{H. L\"u,$^{1,2,6}$ S. K. \"Ozdemir,$^{3,\dagger}$ L.-M. Kuang,$^{1}$, Franco Nori,$^{4,5,\ddagger}$ and H. Jing$^{1,}$}
\email[]{jinghui73@gmail.com}
\email[$^\ddagger$]{fnori@riken.jp}
\email[$^\dagger$]{sko9@psu.edu}

\affiliation{$^1$Key Laboratory of Low-Dimensional Quantum Structures and Quantum Control of Ministry of Education, Department of Physics and Synergetic Innovation Center for Quantum Effects and Applications, Hunan Normal University, Changsha 410081, China}
\affiliation{$^2$Key Laboratory for Quantum Optics, Shanghai Institute of Optics and Fine Mechanics, Chinese Academy of Sciences, Shanghai 201800, China}
\affiliation{$^3$Department of Engineering Science and Mechanics, The Pennsylvania State University, University Park, Pennsylvania 16802, USA}
\affiliation{$^4$CEMS, RIKEN, Saitama 351-0198, Japan}
\affiliation{$^5$Physics Department, The University of Michigan, Ann Arbor, Michigan 48109-1040, USA}
\affiliation{$^6$University of Chinese Academy of Sciences, Beijing 100049, China}

\date{\today}

\begin{abstract}
Intrinsic defects in optomechanical devices are generally viewed to be detrimental for achieving coherent amplification of phonons, and great care has been exercised in fabricating devices and materials with no (or minimal number of) defects. Contrary to this view, here we show that, by surpassing an exceptional point (EP), both the mechanical gain and the phonon number can be enhanced despite increasing defect losses. This counterintuitive effect, well described by an effective non-Hermitian phonon-defect model, provides a mechanical analog of the loss-induced purely-optical lasing. This opens up the way to operating random-defect phonon devices at EPs.
\end{abstract}

\maketitle

\thispagestyle{empty}
\twocolumngrid

\section{Introduction}

Advances in cavity optomechanics (COM) in the past decade have led to many practical applications, such as ultrasensitive motion sensors, quantum transducers, and low-noise phonon devices \cite{AKM2014,Metcalfe2014}. The phonon analog of an optical laser was also achieved in COM \cite{Vahala2010}. Compared to phonon lasers in, e.g., cold ions, superlattices, or electromechanical systems \cite{Vahala2009,Mahboob2013,Kent2010}, COM-based devices feature a continuously tunable gain spectrum to selectively amplify phonon modes, from radio frequency to microwave rates, with an ultralow threshold \cite{Vahala2010,Xiao2017}. This provides a powerful tool to study quantum acoustic effects, e.g., two-mode correlations \cite{Lawall2014}, sub-Poissonian distributions \cite{Painter2015}, and motion squeezing \cite{Wollman2015,Lei2016,Hammerer2015}, which are useful in enhancing the performance of phonon devices in acoustic sensing, imaging, or switching \cite{phonon-rmp,Alu,Yamaguchi2013,comb,capacitor,imaging,computing}.

Very recently, COM devices with balanced gain and loss have also attracted growing interest \cite{Peng2014,Jing2014,Jing2015,Jing2016,2015Chaos,Jing2017,diode,El-Ganainy2016,He2016}. The gain is provided by doping active materials, e.g., rare-earth ions or dyes, into the resonator \cite{Peng2014}. Such systems exhibit non-Hermitian degeneracies known as exceptional points (EPs), where both the eigenvalues and the corresponding eigenfrequencies of the system coalesce. Approaching an EP drastically affects the dynamics of a physical system, leading to many unconventional effects, e.g., loss-induced coherence \cite{Guo2009,Peng2014Science}, invisible sensing \cite{Lin2011,Regensburger2012,Peng2016}, and chiral-mode switch \cite{Doppler2016}. Alternative EP physics has also been explored experimentally in acoustic \cite{Alu2015}, electronic \cite{Kottos2013}, and atomic systems \cite{Xiao2016}, as well as in a COM device \cite{Xu2016}, opening up the way to phononic engineering at EPs.

In this work, we study the emergence of an EP in COM. The EP arises in the phonon-lasing regime, by tuning the loss of intrinsic two-level-system (TLS) defects naturally existing in amorphous materials used in the fabrication of COM devices \cite{Morita1990,Zeller1971,Anderson1972,Phillips1972,Golding1976,Anghel2007,Yu1994,Tanaka2009,Martinis2005,Nori2006,Neeley2008,qubit,Osborn2016}. In a COM system, the role of TLS defects was already studied in the phonon-cooling regime \cite{Ramos2013}, but it has been neglected thus far in the phonon-lasing regime. Counterintuitively, we find that, in the phonon-lasing regime, increasing the defect loss leads to the enhancement of both mechanical gain and emitted phonon number. Unlike similar optical loss-induced effects \cite{Guo2009,Peng2014Science,Osborn2016}, our work provides a route for achieving an EP-enhanced phonon laser without any optical gain. In view of rapid advances in phonon devices \cite{phonon-rmp}, EP optics, and COM with defects \cite{Osborn2016,Tian2011,Ramos2013}, our findings hold the promise of being observed in practical phonon-laser systems with intrinsic defects.

\begin{figure}[ht]
\centering
\includegraphics[width=3.3in]{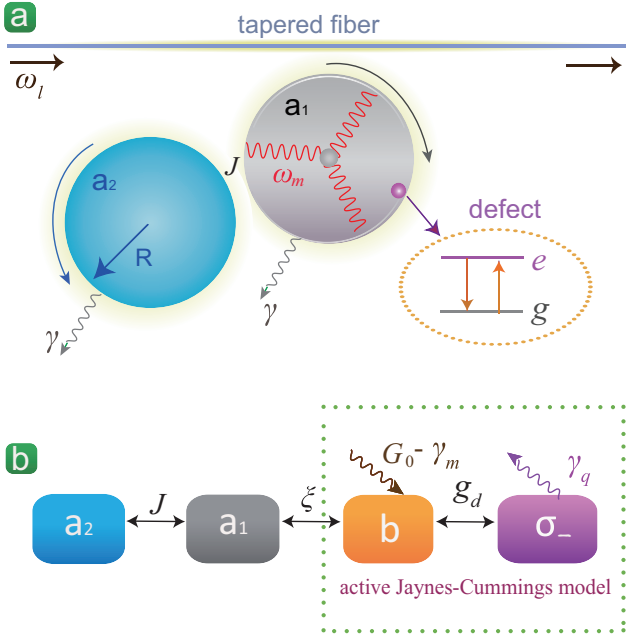}
\caption{Schematic illustrations of the random-defect phonon laser. (a) The top resonator, coupled to a tapered fiber, supports a mechanical mode $b$ \cite{Vahala2010} and contains a defect-induced TLS. (b) The optical mode $a_1$, which is coupled to mode $a_2$ with strength $J$, interacts with $b$, which, in turn, is coupled to the TLS with the strength $g_d$. The TLS decay rate is denoted by $\gamma_q$, and $\gamma_m'=\gamma_m-G_0$ is the effective mechanical damping rate, while $G_0$ is the mechanical gain.}
\label{fig1}
\end{figure}

TLS defects can couple to different modes of a system via different mechanisms, e.g., to superconducting qubits \cite{Martinis2005,Nori2006,Neeley2008,qubit} via electric dipole moments and to phonons via strain forces \cite{Zeller1971,Anderson1972,Phillips1972}. For many years, TLS defects were considered as a main source of loss and decoherence, and as such, techniques have been developed to decrease the number of defects \cite{Zeller1971,Anderson1972,Phillips1972,Golding1976,Anghel2007,Yu1994,Tanaka2009,Martinis2005}. However, recent studies have shown that they can play useful roles in, e.g., TLS quantum memory \cite{Nori2006,Neeley2008}, circuit control \cite{qubit}, and optical lasing \cite{Osborn2016}. In COM systems, a strong TLS-phonon coupling, well described by a Jaynes-Cummings-like model, was utilized to achieve phonon cooling \cite{Tian2011,Ramos2013}. Here we show that phonon lasing can be enhanced by steering lossy defects \cite{Phillips1972,Yu1994}, instead of using any additional loss compensation technique via gain materials. Our work provides a scheme to realize loss-induced phonon lasing in COM systems and to use it for steering phonon devices. The critical point, observed for our full Hermitian system, coincides well with an EP emerging in an effective non-Hermitian TLS-phonon system. Despite its similarity to the loss-induced revival of an optical laser \cite{Peng2014Science}, both the underlying coupling and the critical condition for our COM system, as shown here, are clearly different.

\section{Model and Solutions}

We consider two whispering-gallery-mode resonators (having the same resonance frequency $\omega_c$ and loss rate $\gamma$; see Fig.\,1), one of which supports a radially symmetric mechanical breathing mode with effective mass $m$, frequency $\omega_m$ and damping rate $\gamma_m$. The resonators are made of silica, silicon, or silicon nitride that has intrinsic or artificially doped TLS defects, which can be coupled to the phonon mode via mechanical strain \cite{Tanaka2009,Tian2011,Ramos2013}. The strength of the coupling between the TLS and the mechanical mode, derived from linear elastic solid theory \cite{Ramos2013}, is given as
\begin{equation}
g_d \approx \frac{D_T}{\hbar}\frac{\Delta_0}{\omega_q}S_{\rm zpf},~~~S_{\mathrm{zpf}}=\sqrt{\frac{\hbar\omega_m}{2YV_m}}
\end{equation}
where $\omega_q=\sqrt{\Delta^2_0+\Delta^2_a}$ is the tunable energy difference of the excited and ground states of the TLS \cite{Phillips1972,Tanaka2009}, $\Delta_0$ is the tunnel splitting frequency, $\Delta_a$ is the asymmetry frequency, $S_{\mathrm{zpf}}$ is the zero-point strain-field fluctuation \cite{Ramos2013,Golding1976}, $Y$ is the Young's modulus, and $V_m$ is the mechanical mode volume determined by tensorial strain profiles \cite{Ramos2013}. Note that the mechanical deformation potential $D_T$ can be measured experimentally \cite{Golding1976}, and $\Delta_a$\textemdash and thus $\omega_q$ or $g_d$\textemdash can be tuned using external microwave or electric fields \cite{Osborn2016}. $\delta\omega_q/2\pi\sim1\,$MHz is achievable with a moderate field of about $ 10^3\,\rm V/m$, allowing a TLS-phonon coupling which is strong enough to exceed $\gamma$  \cite{Golding1976,Yu1994}. The TLS-phonon coupling $g_d$ is strong enough to exceed $\gamma$, as previously shown \cite{Ramos2013}. In addition, as shown in Ref.\,\cite{Ramos2013}, the typical number $N_T$ of TLSs that can couple resonantly to a phonon mode, i.e., those within the bandwidth $g_d$ around $\omega_m$, is estimated as $N_T\lesssim 1$ or $N_T\ll 1$, which can be further tuned by e.g., shifting an off-resonant TLS into resonance with the considered phonon mode \cite{Ramos2013}.

In the rotating frame at the pump frequency $\omega_l$, the Hamiltonian of the defect-COM system can be written at the simplest level as
$H = H_0 + H_{\text{int}} + H_{\text{dr}},$ with
\begin{align}
H_0 =\,& -\Delta (a^{\dagger}_{1}a_{1} +a^{\dagger}_{2}a_{2}) + \omega_{m} b^{\dagger}b + \frac{ \omega_{q}}{2}\sigma_{z},\nonumber\\
H_{\text{int}} =\,& J(a^{\dagger}_{1}a_{2} +a^{\dagger}_{2}a_{1})- \xi a^{\dagger}_{1} a_{1}x + g_d (b^{\dagger}\sigma_{-} + \sigma_{+}b),
\end{align}
and $$H_{\text{dr}} =i(\varepsilon_{l}a^{\dagger}_{1} - \varepsilon^{\ast}_l a_{1}),$$
where $a_1$, $a_2$, or $b$ denote the annihilation operators of the optical modes or the mechanical mode, $x=x_0 (b^{\dagger} + b)$ is the mechanical displacement operator, $$\Delta\equiv \omega_l-\omega_c$$ denotes the detuning between the pump laser and the cavity resonance, $\xi=\omega_c/R$ is the COM coupling strength, $R$ is the resonator radius, $x_0=(1/2m\omega_m)^{1/2}$, while $\sigma_z$, $\sigma_{-}$ and $\sigma_{+}$ are the Pauli operators of the TLS defined as $$\sigma_z=|e\rangle\langle e|-|g\rangle\langle g|,~~ \sigma_{-}=|g\rangle\langle e|,~~\sigma_{+}=|e\rangle\langle g|.$$ The pump field amplitude is given by
$$\varepsilon_{l}=(2P_l\gamma/\hbar\omega_l)^{1/2},$$ where $P_l$ is the pump power. The Jaynes-Cummings-like model describes the strain-induced TLS-COM coupling, all details of theoretical derivations of which can be found in previous works on the defect-assisted COM, based on a linear elastic solid-state theory \cite{Tian2011,Ramos2013,Tanaka2009}. The parameter values used in our numerical simulations satisfy the validity condition
$$g_d\ll\omega_q\approx\omega_m$$
of this effective model \cite{Ramos2013}.

\subsection{Supermode picture}

To derive the Hamiltonian in the optical supermode picture, we define the operator $a_{\pm}=(a_1\pm a_2)/\sqrt{2},$ which transforms $H_0$ and $H_{\text{dr}}$ into
\begin{align}	
\mathcal{H}_0 &= \omega_{+} a^{\dagger}_{+}a_{+} + \omega_{-} a^{\dagger}_{-}a_{-} + \omega_{m} b^{\dagger}b + \frac{\omega_q}{2}\sigma_{z}, \nonumber\\ \mathcal{H}_{\text{dr}}&=\frac{i}{\sqrt{2}}\left[\varepsilon_{l}(a^{\dagger}_{+}+a^{\dagger}_{-})-h.c. \right],
\end{align}
with $\omega_{\pm}=-\Delta \pm J.$ Similarly, $H_{\text{int}}$ becomes
\begin{align}
\mathcal{H}_{\text{int}} = &-\frac{\xi x_0}{2}\left[(a^{\dagger}_{+}a_{+}+a^{\dagger}_{-}a_{-})-(a^{\dagger}_{+}a_{-}
+\text{H.c.})(b^{\dagger}+b)\right]\nonumber\\
&+g_d (b^{\dagger}\sigma_{-} + \sigma_{+}b).
\end{align}
In the rotating frame with respect to $\mathcal{H}_0$, we have
\begin{align}
\mathcal{H}_{\text{int}}= &-\frac{\xi x_0}{2}\left(a^{\dagger}_{+}a_{-}be^{i(2J-\omega_m)t}+\text{H.c.}\right)\nonumber\\
&-\frac{\xi x_0}{2}\left(a^{\dagger}_{+}a_{-}b^{\dagger}e^{i(2J+\omega_m)t} +\text{H.c.}\right)
\nonumber\\
&+\frac{\xi x_0}{2}\left(a^{\dagger}_{+}a_{+} + a^{\dagger}_{-}a_{-}\right)\left(b^{\dagger}e^{i\omega_m t} + \text{H.c.}\right)\nonumber\\
&+ g_d \left[b^{\dagger}\sigma_{-}e^{i(\omega_m-\omega_q)t}+\text{H.c.}\right].
\end{align}
Considering the rotating-wave approximation $$2J+\omega_m,\omega_m\gg |2J-\omega_m|,|\omega_q-\omega_m|,$$ we have
\begin{equation}
\mathcal{H}_{\text{int}}= - \frac{\xi x_0}{2} (a_{+}^{\dagger}a_{-}b + b^{\dagger}a_{+}a_{-}^{\dagger}) + g_d (b^{\dagger}\sigma_{-} + \sigma_{+}b).
\end{equation}
The first term describes the phonon-mediated transition between optical supermodes, and the second term describes the coupling between the phonon and the TLS defect. Thus, in the supermode picture, the optomechanical coupling is transformed into an effective coupling describing defect-assisted phonon lasing. The TLS can be excited by absorbing a phonon generated from the transition between the upper optical supermode and the lower one, and as such it can strongly modify the behavior of the phonon lasing.

The Heisenberg equations of motion of the system can then be written as
\begin{align}
\dot{a}_{+}&=(-i\omega_{+}-\gamma)a_{+} +\frac{i\xi x_0}{2}a_{-}b + \frac{\varepsilon_l}{\sqrt{2}} + \sqrt{\gamma}a_{\rm in},\nonumber\\
\dot{a}_{-}&=(-i\omega_{-}-\gamma)a_{-} +\frac{i\xi x_0}{2}a_{+}b^{\dagger}+ \frac{\varepsilon_l}{\sqrt{2}}+\sqrt{\gamma}a_{\rm in},\nonumber\\
\dot{b}&=(-i\omega_{m} - \gamma_{m})b + \frac{i\xi x_0}{2} a^{\dagger}_{+}a_{-} - ig_d\sigma_{-} + \sqrt{2\gamma_{m}}b_{\rm in}\nonumber,\\
\dot{\sigma}_{-}&=(-i\omega_{q}- \gamma_{q})\sigma_{-} + ig_d b \sigma_{z} + \sqrt{2\gamma_{q}}\Gamma_{-}\nonumber,\\
\dot{\sigma}_{z}&=-2\gamma_{q}(\sigma_{z}+1)-2ig_d(\sigma_{+}b-b^{\dagger}\sigma_{-})+\sqrt{2\gamma_{q}}\Gamma_{z}.
\label{Eq:Heisenberg0}
\end{align}
Here $a_{\rm in}$, $b_{\rm in}$, $\Gamma_{-}$, and $\Gamma_z$ denote environmental noises corresponding to the operators  $a$, $b$, $\sigma_{-}$ and $\sigma_z$. We assume that the mean values of these noise operators are zero, i.e.
$$\left\langle a_{\rm in} \right\rangle=\left\langle b_{\rm    in}\right\rangle=\left\langle\Gamma_{-}\right\rangle=\left\langle\Gamma_z\right\rangle=0.$$ The fluctuations are
small and we neglect the noise operators in our numerical calculations. Then, the defect-assisted mechanical gain and the threshold power of the phonon lasing can be obtained.

In the supermode picture, a crucial term describing the phonon-lasing process can be resonantly chosen from the Hamiltonian, under the rotating-wave approximation \cite{Vahala2010,HW2014} (for $J\sim \omega_m/2$, $\omega_q\sim \omega_m$). The resonance $\omega_m=\omega_q$ can be achieved by using a moderate field of about $10^3\,\rm V/m$, allowing a shift of $\delta\omega_q/2\pi\sim1\,$MHz \cite{Ramos2013,Golding1976,Yu1994}. With the supermode operators $p=a^{\dagger}_{-}a_{+}$, $a_{\pm}=(a_1\pm a_2)/\sqrt{2}$, the reduced interaction Hamiltonian is given by
\begin{align}
\mathcal{H}_{\text{int}}= - \frac{\xi x_0}{2} (p^\dag b + b^{\dagger}p) + g_d (b^{\dagger}\sigma_{-} + \sigma_{+}b).
\end{align}
The resulting Heisenberg equations of motion are
\begin{align}
\dot{p}&=(-2iJ-2\gamma)p-\frac{i\xi x_0}{2}\delta n b + \frac{1}{\sqrt{2}}(\varepsilon^{\ast}_{l}a_{+}+\varepsilon_{l}a^{\dagger}_{-}),\nonumber\\
\dot{b}&=(-i\omega_{m} - \gamma_{m})b + \frac{i\xi x_0}{2}p - ig_d \sigma_{-} \nonumber,\\
\dot{\sigma}_{-}&=(-i\omega_{q}- \gamma_{q})\sigma_{-} + ig_d b \sigma_{z} \nonumber,\\
\dot{\sigma}_{z}&=-2\gamma_{q}(\sigma_{z}+1) - 2ig_d(\sigma_{+}b - b^{\dagger}\sigma_{-}),
\label{Eq:Heisenberg}
\end{align}
where $p=a^{\dagger}_{-}a_{+}$, and $\delta n=a^{\dagger}_{+}a_{+}-a^{\dagger}_{-}a_{-}$ denotes the population inversion. The noise terms are negligible with a strong driving field. The steady-state values of the system can be obtained by setting $\partial p/\partial t=0,\partial\sigma_-/\partial t=0$, and $\partial a_{\pm}/\partial t=0$, with $\gamma,\gamma_q\gg\gamma_m$, which leads to
\begin{align}
\label{eq:eliminate}
p&=\frac{1}{i(2J-\omega_m)+2\gamma}\left[\frac{1}{\sqrt{2}}(\varepsilon^{\ast}_{l}a_{+}+\varepsilon_{l}a^{\dagger}_{-})  -\frac{i\xi x_0}{2}\delta n b \right],\nonumber\\
\sigma_{-}&=-\frac{g_d(\omega_q-\omega_m) + ig_d\gamma_q}{\gamma^2_q+(\omega_q-\omega_m)^2+2g_d^2 n_b}b,\nonumber\\
a_{+}&=\frac{\varepsilon_l(2i\omega_{-} + 2\gamma + i\xi x_0 b)}
{2\sqrt{2}\alpha-i4\sqrt{2}\gamma\Delta},\nonumber\\
a_{-}&=\frac{\varepsilon_l(2i\omega_{+} + 2\gamma + i\xi x_0 b^{\dagger})}
{2\sqrt{2}\alpha-i4\sqrt{2}\gamma\Delta},
\end{align}
where $n_b$ denotes the expectation value of the phonon number and
\begin{equation}
\omega_{\pm}=-\Delta\pm J,~~~\alpha=J^2+\gamma^2-\Delta^2 +\frac{\xi^2 x^2_0}{4}n_b.\nonumber
\end{equation}
Substituting these values into the equation of the mechanical mode results in
\begin{equation}
\dot{b}=\left(-i\omega_m + i\omega^{\prime} + G - \gamma_m\right)b  + C,
\end{equation}
where
\begin{align}
\omega^{\prime}=&\frac{g_d^2(\omega_q-\omega_m)}{\gamma^2_q+(\omega_q-\omega_m)^2+2g_d^2n_b} - \frac{\xi^2x^2_0(2J-\omega_m)}{16\gamma^2+4(2J-\omega_m)^2}\nonumber\\
&-\frac{\xi^2x^2_0\Delta|\varepsilon_l|^2}{[2(2J-\omega_m)^2+8\gamma^2](\alpha^2+4\Delta^2\gamma^2)}, \nonumber\\
C=&\frac{i|\varepsilon_l|^2\xi x_0}{2i(2J-\omega_m)+4\gamma}\cdot\frac{(\gamma-iJ)\alpha+2\Delta^2\gamma}{\alpha^2 + 4\Delta^2\gamma^2}, \nonumber
\end{align}
and $$\alpha=J^2+\gamma^2-\Delta^2 +{\xi^2 x^2_0}n_b/{4}.$$ The mechanical gain is then $G=G_0+G_d$, with
\begin{align}
G_0=&\frac{\xi^2 x^2_0\gamma}{2(2J-\omega_m)^2+8\gamma^2} \left(\delta n- \frac{\Delta(2J-\omega_m) |\varepsilon_l|^2}{\alpha^2 + 4\Delta^2\gamma^2}\right),\nonumber\\
G_d=&-\frac{g_d^2\gamma_q}{\gamma^2_q+(\omega_q-\omega_m)^2+2g_d^2n_b}.
\end{align}
The role of lossy defects in mechanical amplification, described by $G_d$, has not been reported previously. From the condition $G=\gamma_m$ and $P_{\mathrm{th}}\approx\hbar(\omega_c+J)\gamma\delta n$ \cite{Vahala2010}, the threshold power $$P_\mathrm{th}=P_{\text{th,0}}+P_{\text{th},d}$$ is found as
\begin{align}
P_{\text{th,0}} &= \frac{2\hbar\left[(2J-\omega_m)^2+4\gamma^2\right](\omega_c+J)\gamma_m}{(\xi x_0)^{2}} \nonumber \\
&+\frac{\hbar\Delta(2J-\omega_m)(\omega_c+J) |\varepsilon_l|^2}{\lambda^2+4\Delta^2\gamma^2}, \nonumber   \\
P_{\text{th},d} &= \frac{2\hbar g^2_d\gamma_q(\omega_c+J)\left[(2J-\omega_m)^2+4\gamma^2\right]}{\xi^2 x^2_0 \left[\gamma^2_q+(\omega_m-\omega_q)^2 + 2g_d^2n_b\right]}.
\end{align}
Clearly, the presence of defects strongly alters $G$ and $P_{\text{th}}$, even when $\Delta=0$. In the following, we first present the full numerical results, and then, to understand the observed counterintuitive effect, we introduce a reduced non-Hermitian TLS-phonon model. A comparative analysis of the full and reduced models then helps to establish the relation between the turning points of the former with the EPs emerging in the latter.

\begin{figure}[ht]
\centering
\includegraphics[width=3in]{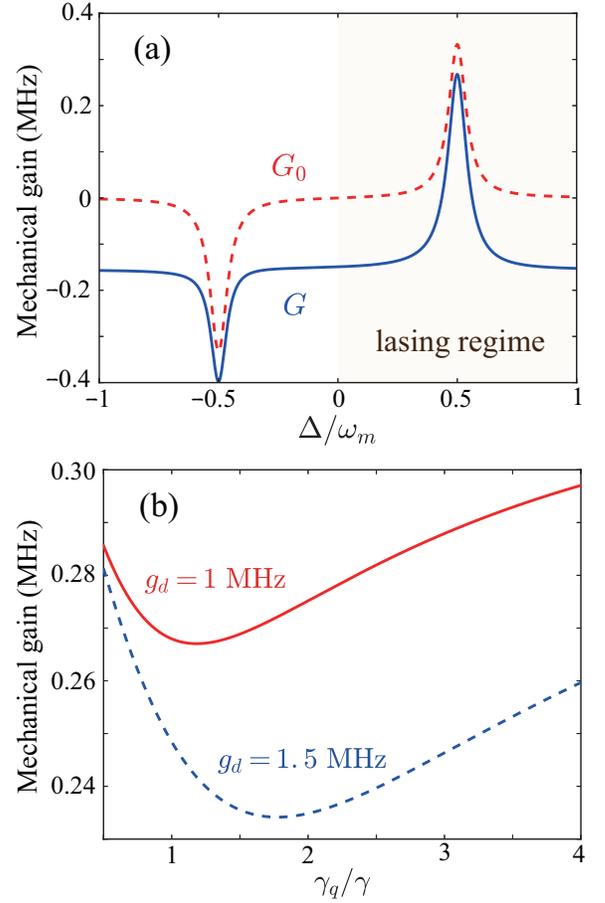}
\caption{(a) The mechanical gains $G_0$ (without defect) and $G$ (with defect) as a function of the optical detuning $\Delta$. (b) $G$ as a function of the defect loss $\gamma_q$. Here, for simplicity, we use $\gamma_q/\gamma=1$, $g_d=1\,\mathrm{MHz}$ in (a), $\Delta=0.5\omega_m$ in (b), and $J=0.5\omega_m$ and $P_l=10\,\mu\mathrm{W}$ in both (a) and (b).}
\label{fig2}
\end{figure}

\section{Numerical Results and discussions}

\subsection{The full system: Numerical results}

\begin{figure}[ht]
\centering
\includegraphics[width=3in]{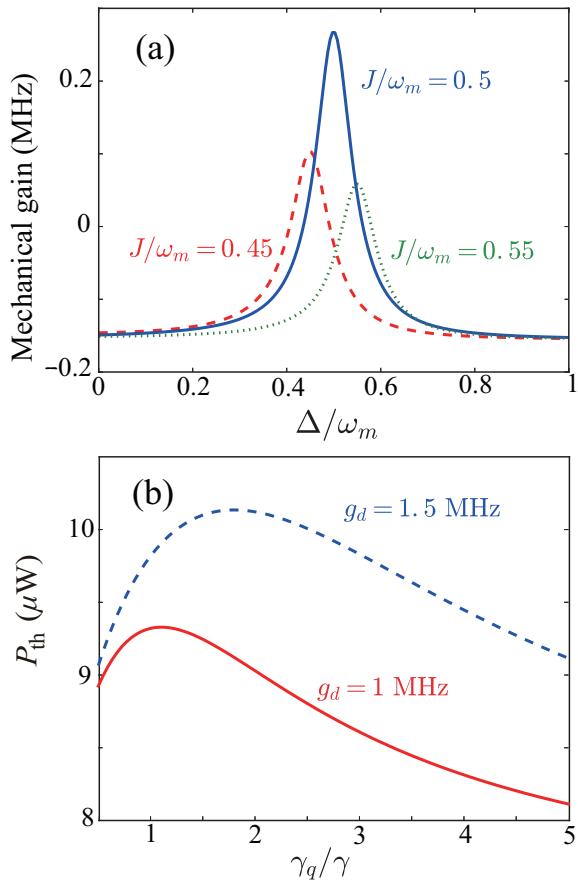}
\caption{(a) Mechanical gain $G$ in a defect-assisted phonon laser versus the pump-cavity detuning $\Delta$. (b) The threshold power $P_{\mathrm{th}}$ of the defect-assisted phonon laser versus the TLS decay rate $\gamma_q$. The parameters used here are (a) $\gamma_q=\gamma$, $g_d=1\,\mathrm{MHz}$ and (b) $J=0.5\omega_m$, $\Delta=0.5\omega_m$, and $\omega_q=\omega_m$ and $P_l=10\,\mu\mathrm{W}$ in both (a) and (b).}
\label{fig3}
\end{figure}

Figure \,\ref{fig2}(a) shows the mechanical gain, $G_0$ and $G$, as a function of the optical detuning $\Delta$, using experimentally accessible values \cite{Vahala2010,Peng2014}, i.e. $R= 34.5\,\mu$m, $m$ = 50\,ng, $\omega_c = 193$\,THz, $\omega_m = 2 \pi \times 23.4$\,MHz, $\gamma = 6.43$\,MHz, and $\gamma_m = 0.24$\,MHz. In the cooling regime (with $\Delta<0$), $G$ is negative and can be enhanced by defects \cite{Ramos2013}. In the lasing regime (with $\Delta>0$), the positive $G$ is also strongly affected by defects. Note that the
simplified condition $\gamma_q/\gamma=1$ used in Fig.\,\ref{fig2}(a) is experimentally accessible, since $\gamma_q$ is typically $0.1-5\,$MHz \cite{Ustinov2010} and can be further enhanced by using external fields (or amorphous oxide layers) \cite{Tanaka2009}. Clearly, the defect-induced reduction in $G$ is minimized at $\Delta/\omega_m\sim 0.5$, and as Fig.\,\ref{fig3}(a) shows, the maximum phonon lasing occurs at $\Delta/\omega_m\sim 0.5$, $J/\omega_m\sim 0.5$.

We stress that the TLS defects naturally and inevitably exist in all solid-state materials and introduce detrimental losses in optomechanical systems. Therefore, the ideal mechanical gain (in the absence of any defect) $G_0$ can never be achieved in a practical device. In order to obtain $G\rightarrow G_0$, the intuitive way is to minimize, if not eliminate, the detrimental effects of TLS defects by preparing better and purer materials with no or minimal number of defects.
In contrary to this view, we find that this can also be achieved by increasing the losses induced by TLS defects (e.g. by controlling the dissipation of existing defects or by introducing more defects that are coupled to the mechanical mode). 
As shown in Fig.\,\ref{fig2}(b), a turning point appears for $G$ as the TLS loss is increased: $G$ first decreases with increasing TLS loss, until a critical value of $\gamma_q$. When this value is exceeded, more loss leads to an increasing $mechanical$ gain, tending to the limit value $G_0$ as we have numerically confirmed. Consequently, the phonon-lasing threshold power $P_{\mathrm{th}}$ first increases and then decreases again with more loss, as shown in Fig.\,3(b).
This counterintuitive effect, emerging $only$ in the mechanical-amplifying regime, has not been reported previously. Despite the similarity to loss-induced purely optical-lasing revival \cite{Peng2014Science,Guo2009}, the underlying coupling and the critical condition of our hybrid COM system are clearly different.

\begin{figure}[ht]
\centering
\includegraphics[width=3in]{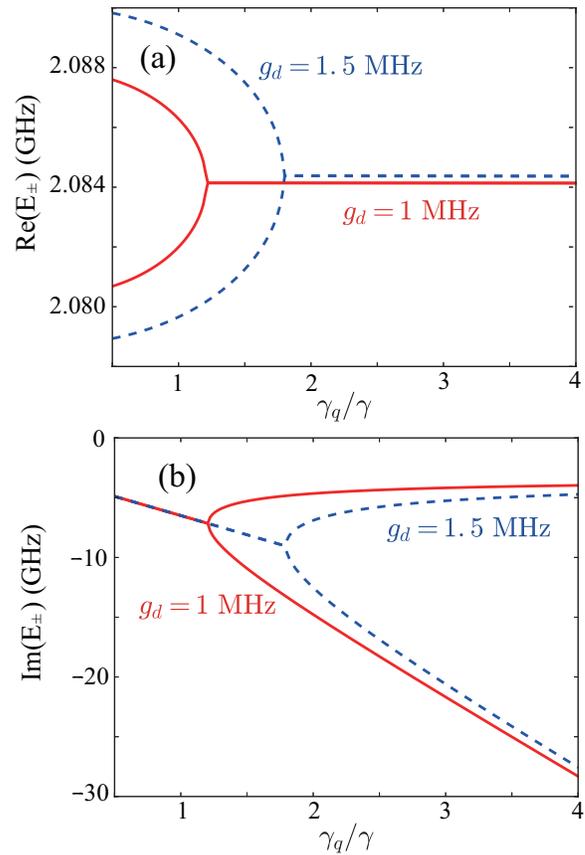}
\caption{Supermode spectrum of the reduced system, i.e., the (a) real and (b) imaginary parts of the eigenvalues of $\mathcal{H}_\mathrm{eff}$, with the experimentally accessible values of the tunable parameters $J=0.5\omega_m$, $\gamma_q/\gamma=1$, and $P_l=7\,\mu\mathrm{W}$ (the threshold value as measured in the experiment \cite{Vahala2010}).}
\label{fig4}
\end{figure}

\subsection{Active Jaynes-Cummings model}

\begin{figure}[ht]
\centering
\includegraphics[width=3.3in]{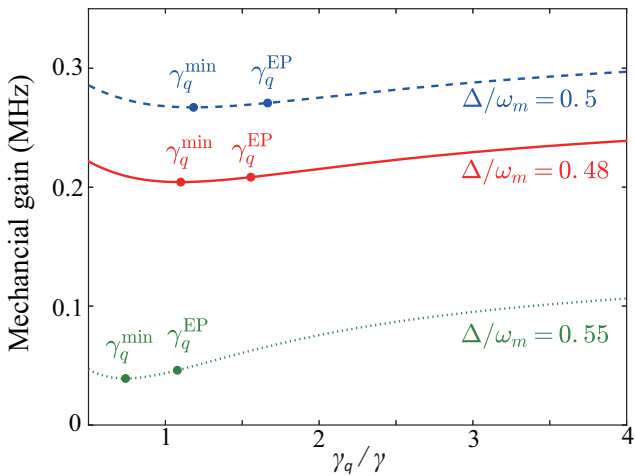}
\caption{Mechanical gain $G$ as a function of the TLS loss rate $\gamma_q$, for different values of $\Delta$. The parameters used here are $J=0.5\omega_m$, $\omega_q/\omega_m=1$, and $P_l=10\,\mu\mathrm{W}$.}
\label{fig5}
\end{figure}

\begin{figure}[ht]
\centering
\includegraphics[width=3.3in]{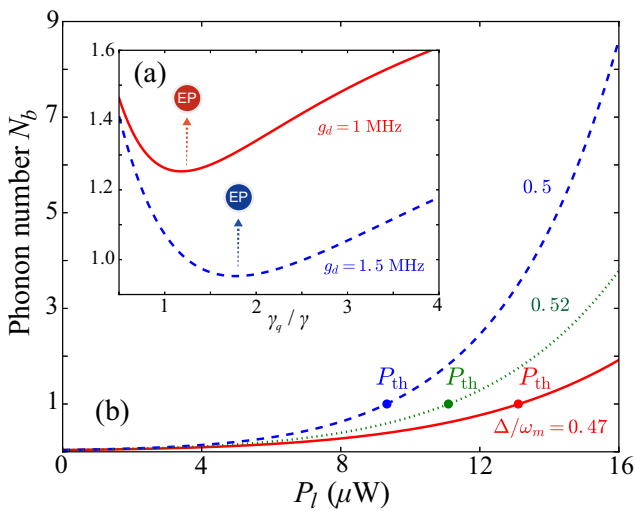}
\caption{Stimulated emitted phonon number. The phonon number $N_b$ versus (a) the pump power $P_l$ and (b) the damping rate $\gamma_q$, with $\omega_q/\omega_m=1$. Also, we take the achievable values (a) $P_l=10\,\mu\mathrm{W}$ and (b) $\gamma_q/\gamma=1$, $g_d=1\,\mathrm{MHz}$.}
\label{fig6}
\end{figure}

To intuitively understand the turning-point feature, as numerically revealed above, we resort to a reduced model with only the $active$ phonon mode and the lossy defects, i.e.,
\begin{equation}
\mathcal{H}_\mathrm{eff}=(\omega_m-i\gamma_m^{\prime}) b^{\dagger}b + (\omega_q-i\gamma_q) \sigma_{+}\sigma_{-}+g_d(b^{\dagger}\sigma_{+}+\sigma_{-}b),\nonumber
\end{equation}
with the effective damping
$$
\gamma^{\prime}_m=\gamma_m-G_0.$$
We note that in a recent experiment\cite{Xiao2016}, a similar route was adopted for achieving a non-Hermitian atomic system, where, by starting from a Hermitian Hamiltonian describing full atom-light interactions, an effective non-Hermitian model was deduced for atomic excitations (see also Ref.\,\cite{GWunner2013}). Choosing two basis states, $|n_b,g\rangle$ and $|n_b-1,e\rangle$, to diagonalize $\mathcal{H}_\mathrm{eff}$ leads to the eigenvalues
\begin{align}
E_{\pm}=&\left(n_b-\frac{1}{2}\right)\omega_m +\frac{\omega_q}{2}-\frac{i}{2}\left[(2n_b-1)\gamma^{\prime}_m+\gamma_q\right]\nonumber\\
&\pm\frac{1}{2}\sqrt{4n_b g^2_d+\left[\omega_q-\omega_m-i(\gamma_q-\gamma^{\prime}_m) \right]^2 }.
\end{align}
The supermode spectrum of these eigenvalues is shown in Fig.\,\ref{fig4}(a) and \ref{fig4}(b), where an EP is seen at the position close to the turning points in Fig.\,\ref{fig2}(b) and Fig.\,\ref{fig3}(b). This EP, labelled by the critical value $\gamma_q^{\rm EP}$, characterizes the transition between two distinct phases of the hybrid TLS-phonon system \cite{Peng2014Science,Bender1998,Bender1999,Bender2007,Bagarello2015}:

($\rm i$) For $\gamma_q\leq\gamma_q^\mathrm{EP}$, the supermodes are almost equally distributed between the phonons and the defects, and the active phonon mode partially or completely compensates for loss induced by the defects. Consequently, as $\gamma_q$ is increased, the system has less net mechanical gain.

($\rm ii$) For $\gamma_q>\gamma_q^\mathrm{EP}$, the supermodes become increasingly localized such that one dominantly resides in the phonon mode and the other in the defects. Hence with increasing $\gamma_q$, the supermode which is dominant in the defects experiences more loss, while the supermode which is dominant in the phonon mode experiences less loss (i.e., increased mechanical gain).

For the special case $\omega_q/\omega_m=1$ \cite{Ramos2013}, the EP emerges at $$\gamma_q^{\rm EP}=\gamma'_m+2\sqrt{n_b}g_d.$$ while, when $\partial G/\partial\gamma_q=0$, the turning point of $G$ is obtained at
\begin{equation}
\gamma^{\mathrm{min}}_q=\sqrt{2n_b}\,g_d.
\end{equation}
The slight shift of the turning point from the exact EP position is due to the fact that $\gamma^{\mathrm{min}}_q$ depends on $\Delta$, while the EP does not. A comparison of the turning points and the EPs for different values of the optical detuning is given in Fig.\,\ref{fig5}. We note that the slight shift of the turning point from the exact EP position was also observed in a purely-optical system (see Ref.\,\cite{Peng2014Science}).
We also note that the EP of this TLS-phonon system is reminiscent of that observed recently in a Jaynes-Cummings system with a single atom trapped in a high-$Q$ cavity (by using, however, a different method of tuning the atom-cavity coupling) \cite{Choi2010}.

Finally, Fig.\,\ref{fig6} shows the phonon number $$N_b=\exp\left[2(G-\gamma_m)/\gamma_m\right],$$ as a function of the defect loss and the pump power. Features similar to those observed for the mechanical gain also appear for $N_b$, i.e. more loss leads to the suppression of $N_b$ for $\gamma_q\leq\gamma_q^\mathrm{min}$, but $N_b$ is enhanced  with more loss for $\gamma_q>\gamma_q^\mathrm{min}$. The turning point of $N_b$ is in exact correspondence with that of the mechanical gain, as shown in Fig.\,2(b), or the threshold power in Fig.\,3(b). Figure\,6(b) shows that $N_b$ is strongly dependent on $\Delta$, and the optimized condition $\Delta/\omega_m=0.5$, as in the case without defects, still holds in the presence of TLS defects.

\section{Conclusion}

In this work, we study the counterintuitive role of defects in the phonon-lasing process. We find that the exact evolutions of the mechanical gain and the threshold power exhibit a turning point as the loss is increased. This effect is closely related to the emergence of an EP in an effective non-Hermitian TLS-phonon system. When exceeding the EP, more TLS loss leads to an enhanced mechanical gain, along with a lowered threshold for the phonon laser. This indicates that the detrimental effects of intrinsic lossy defects (naturally existing in solid-state materials) in phonon lasing can be minimized.
This sheds a different light not only on EP physics and optomechanics but also on practical control of random-defect phonon devices.

We note that the COM-based phonon laser has already been experimentally realized \cite{Vahala2010,Xiao2017}, and the effect of inevitably existing defects in the COM device was also studied in the phonon-cooling regime \cite{Ramos2013}. Our work extends the COM-TLS structures to the mechanical-amplifying regime and reveals the emergence of a loss-induced EP. 
We establish the relation between EP, TLS loss, and mechanical amplification, which has not been studied
before.
We also note that, besides material strain \cite{qubit,Ramos2013,Wang2015}, the TLS energy splitting and damping rate can be controlled by external electric fields \cite{Osborn2016}. This opens the way for electrically tuned phonon lasing. Finally, we remark that the optical effect of defects can be incorporated into the optical decay rate, and the mechanical strain only induces the phonon-TLS coupling (not any additional optical effects, see also Ref.\,\cite{Ramos2013}). In our future works, we will consider placing a nanotip near the optical resonator \cite{Peng2014Science} to study the interplay of the interplay of the loss-induced optical EP \cite{Peng2014Science} and the TLS-phonon EP, or placing an atom in the system \cite{Kimble2005} to study the interplay of the atom-photon coupling and the TLS-phonon coupling. It will be also of interesting to study COM squeezing \cite{Hu1996,Hu1997} or sensing \cite{comb,Huang2017} in the presence of TLS-phonon EPs.

\section*{ACKNOWLEDGMENTS}

L.-M.K. is supported by the 973 Program under Grant No. 2013CB921804 and the National Natural Science Foundation of China under Grants No. 11375060 and No. 11434011. H.J. is supported by the National Natural Science Foundation of China under Grants No. 11474087 and No. 11774086. S.K.\"O. is supported by ARO Grant No. W911NF-16-1-0339 and the Pennsylvania State University Materials Research Institute. F.N. is supported by the RIKEN iTHES Project, the MURI Center for Dynamic Magneto-Optics via the AFOSR award number FA9550-14-1-0040, the IMPACT program of JST, CREST No. JPMJCR1676, a JSPS Grant-in-Aid for Scientific Research (A), and a grant from the Sir John Templeton Foundation.

\end{document}